 \documentclass[preprint2]{aastex}

\slugcomment{To appear in The Astrophysical Journal}

\shorttitle{Detecting $O_2$ in an Earth-like Planet}
\shortauthors{Rodler \& L\'opez-Morales}

\begin{document}


\title{Feasibility Studies for the Detection of $O_2$ in an Earth-like Exoplanet}


\author{Florian Rodler}
\affil{Harvard-Smithsonian Center for Astrophysics, Cambridge, MA 02138, USA\\}
\affil{Institut de Cienci\`{e}s de l'Espai (CSIC-IEEC), Campus UAB, Fac. Ciencies, C5 p2, 08193 Barcelona, Spain }
\and
\author{Mercedes L\'{o}pez-Morales}
\affil{Harvard-Smithsonian Center for Astrophysics, Cambridge, MA 02138, USA\\}




\begin{abstract}

We present the results of simulations on the detectability of $O_2$ in the atmosphere of Earth twins around nearby low mass stars using high resolution transmission spectroscopy.
We explore such detectability with each of the three upcoming Extremely Large Telescopes (ELTs), i.e.~GMT, TMT and E-ELT, and high resolution spectrographs, assuming such instruments will be available in all ELTs. With these simulations we extend previous studies by taking into account atmospheric refraction in the transmission spectrum of the exo-Earth and observational white and red noise contributions. Our studies reveal that the number of transits necessary to detect the $O_2$ in the atmosphere of an Earth twin around M-dwarfs is by far higher than the number of transits estimated by Snellen et al. (2013).  In addition, our simulations show that, when accounting for typical noise levels associated to observations in the optical and near-infrared,  the $O_2$ A-band at 760 nm is more favorable to detect the exoplanetary signal than the $O_2$ band at 1268 nm for all the spectral types, except M9V. We conclude that, unless unpredicted instrumental limitations arise,  the implementation of pre-slit optics such as image slicers appear to be key to significantly improve the yield of this particular science case. However, even in the most optimistic cases, we conclude that the detection of $O_2$ in the atmosphere of an Earth twin will be only feasible with the ELTs if the planet is orbiting a bright close-by (d $\le$ 8 pc) M-dwarf with a spectral type later than M3. 
\end{abstract}

 

\section{Introduction} \label{S1}

The path to finding exoplanets either inhabited or amenable 
for life as we know it on Earth includes the detection in their atmospheres of chemical compounds
such as $H_2O$, $CO_2$, $CH_4$,  and $O_3$, known as biomarkers  (Schindler et al. 2000,
Pavlov et al. 2000, Kaltenegger \& Traub 2007, Traub et al. 2008). Another important biomarker is gaseous
oxygen, $O_2$, detected by Sagan et al. (1993) while analyzing the spectrum of Earth observed 
by the Galileo probe. In that spectrum $O_2$ appeared to be, together  with $CH_4$ in strong thermodynamical disequilibrium, 
the strongest indicator of life on Earth. 
$O_2$ has since drawn significant attention because it 
provides a strong indication of the presence of oxygen-producing forms of life and 
because it has several spectral absorption bands at visible and near-infrared wavelengths, which can be detected using 
ground-based telescopes.  The most prominent of those bands are the so-called {\it $O_2$~A-band} around 760 nm,
and the $O_2$~band at 1268 nm, with the A-band being the strongest with 55 strong lines.

Several studies have now investigated the detectability of $O_2$ for an Earth-like exoplanet in the habitable zone
of its host star. The first of those studies was published by Schneider (1994) who analytically calculated the 
detectability of the $O_2$~A-band by estimating the photometric depth of the absorption 
produced by the full band, which has a width of $\sim$ 3 nm. Computing the ratio of the projected surface areas of the planet's atmosphere 
and the disk of the star during a transit, this study concluded that it was possible to detect the $O_2$ A-band with 
a 2.4 meter space-bound telescope, assuming the planet orbits a solar type star at a distance of 2 parsecs or a 0.3 $R_{\sun}$ star at 10 parsecs.  

Webb \& Wormleaton (2001) refined the Schneider (1994) analysis by using empirical observations of the Earth's $O_2$ 
A-band to simulate the transmission spectrum of the exoplanet and by superimposing the resultant spectrum with the spectrum of the host star.
Their analysis concluded that it is possible to detect the planetary $O_2$ A-band for host stars with R $<$ 0.3 $R_{\sun}$, although 
detectability with current 8-meter class telescopes would be limited to M dwarfs with $m_v \sim$ 10. This study also proposed 
that a way to confirm that the $O_2$ lines originated in the atmosphere of the exoplanet is that they should be offset from the Earth's own $O_2$
lines by an amount equal to the peculiar velocity of the host star\footnote{References in the literature trace this technique back to an article 
by Lowell (1905), where he proposes that oxygen lines in the atmosphere of Mars could be detected by their Doppler shift with respect to the corresponding telluric lines.}. However, they did not include the telluric $O_2$ A-band in their analysis.

A few years later, Kaltenegger \& Traub (2009) generated simulations using the Earth's atmosphere as proxy to estimate the detectability of biomarkers
during the transit of an Earth twin around a Sun-like star and also around M-dwarfs, for a 6.5-meter telescope in space. Their main conclusion was
that the signal-to-noise ratio (S/N) of a single transit would not be enough to detect any spectral features in the atmosphere of an exo-Earth orbiting
in the habitable zone of those stars. Instead, multiple transits would be needed, with an estimate of 6 to 192 transits per year for M0V to M9V stars. They 
also concluded that the spectral features of different molecules in the atmosphere of an exo-Earth around an M star will be easier to detect in the infrared
than in the optical, because the stellar flux peaks in the infrared.

Just recently, Snellen et al. (2013) extended previous studies by including the effect of the telluric 
$O_2$ A-band and applying the latest techniques to detect the planet's Doppler shift from the ground (e.g. Brogi et al. 2012, Rodler et al. 2012). 
Their simulations show that it will be possible to detect the $O_2$ A-band in the atmosphere of an Earth twin if a
spectrograph  with a spectral resolving power of $R = 100,000$ is  installed on the next generation of Extremely Large Telescopes (ELTs). In particular, they concluded that the 39-meter 
E-ELT, will be able to detect the $O_2$ A-band transmission signal of a planet around an M5V
star by combining the observations of a few dozen transits. In addition, they remarked that in the case of stars smaller than M7V the $O_2$ might be
easier to detect in the $O_2$band at 1268 nm since those stars will be brighter at near-infrared wavelengths than in the visible.

In this paper we further refine the Snellen et al. (2013) simulations by 1) accounting for atmospheric refraction effects in the exoplanet's 
transmission spectrum and 2) including the effect of random and correlated (atmospheric and instrumental) observational noise. In addition, we extend previous studies by exploring 
the expected performance of instruments for the three upcoming ELTs, exploring a range of spectral resolutions, and performing simulations for a
range of stellar peculiar velocities and spectral types. Finally, we explore in detail the feasibility of the observations both in $O_2$ A-band and the 
$O_2$ band at 1268 nm.

In sections 2 and 3 we describe the implementation and results of our simulations. A discussion of those 
results and the optimal observational set-up is given in section~4.




\section{Simulations}

\subsection{Ingredients}

The first step to perform these $O_2$ detection feasibility studies is to generate a set of hypothetical exo-Earth transiting planet configurations
and observational setups. The parameters necessary for the simulations are 1) a model transmission
spectrum of the Earth-like planet (which we assume to be an Earth twin, including atmospheric conditions and chemistry), 2) a model transmission spectrum of Earth, 3) a model of the
spectral emission of the star, including assumptions about the relative velocity of the star with respect to Earth, 4) a telescope+instrument setup
configuration, and 5) observational noise models.  

\subsubsection{Exoplanet Model Transmission Spectrum}

The model atmospheric transmission spectrum of the exo-Earth was calculated adopting a line-by-line radiative transfer model (LBLRTM) code based of the 
FASCODE algorithm (Clough et al. 2005)\footnote{LBLRTM is available as fortran source code and runs on various platforms. The source code and manuals are available under {\tt http://rtweb.aer.com/lblrtm\_description.html}}. As molecular database for oxygen models we adopted HITRAN (Rothman et al. 2009).

In those calculations we accounted for the effect of refraction in the transmission spectrum of the exo-Earth. As described in  Garcia-Mu\~noz et al. (2012), the transmission spectrum of an Earth twin observed transiting its host star will differ from the Earth's transmission spectrum because of refraction. Refraction is the effect by which light crossing the planet's atmosphere near its surface gets deflected by an angle that depends on the properties of the atmosphere itself, the size of the star, and the orbital distance of the planet. This effect causes the atmospheric layers close to the planet's surface to be invisible for a remote observer. As a consequence, any chemical species at heights lower than that limit can not be observed. For species distributed over a large range of atmospheric heights, as is the case of $O_2$,  the depth of their absorption lines will be altered.  For e.~g., in the particular case of the Earth-Sun system atmospheric layers at heights lower than 12-14 km will be invisible to a remote observer seeing the system transit. In the case of smaller stars, i.e. M-dwarfs,  where both the radius of the star and the orbital distance of an Earth-like planet in the habitable zone of the system are smaller, only atmospheric altitudes lower than about 5 km appear invisible.

In this work we focus on Earth twins orbiting around M-dwarf stars (see section 2.1.3). Therefore, to simulate this refraction effect we only integrate the model transmission spectrum of the exo-Earth at atmospheric layers between 5 km to 85 km from the surface. 85 km marks the end of the Earth's mesosphere, with over 99$\%$ of the atmospheric mass contained below that atmospheric height (Lutgens \& Tarbuck 1995). Therefore, the contribution of the atmospheric layers above that height to the transmission spectrum of the planet can be considered minuscule. 

The resulting model transmission spectrum of the exo-Earth around the $O_2$ A-band as well as the $O_2$ band at 1268 nm are illustrated in the top panels Figures 1 and 2.

\begin{figure}\label{F1}
\includegraphics[scale=0.33, angle=-90]{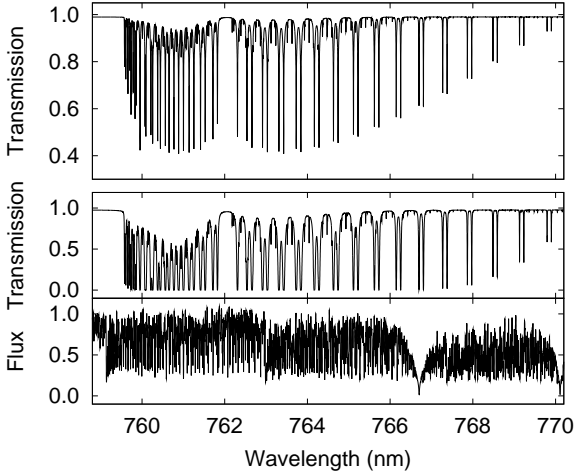}
\caption{Top panel: transmission spectrum of the atmosphere of an Earth-like planet around 760~nm. Middle panel: telluric spectrum of our atmosphere for a zenith distance of $30^\circ$ (airmass $X=1.3$). Bottom: PHOENIX model spectrum of an M4V star with a surface temperature of $T_{\rm eff}=3000$~K, $\log g = 4.5$~dex and solar abundance. All spectra are shown at a spectral resolution of $R=100,000$.}
\end{figure}

\begin{figure}\label{F2}
\includegraphics[angle=-90,scale=0.32]{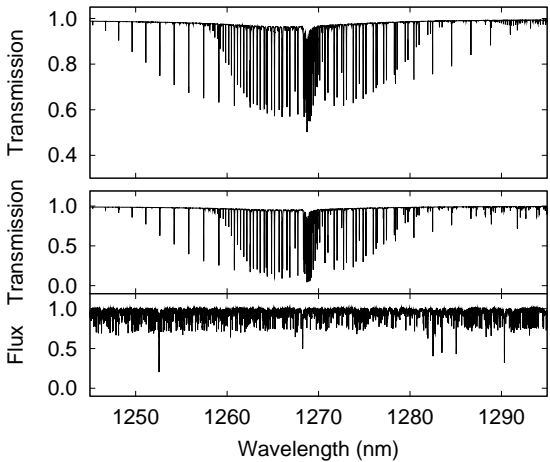}
\caption{Same as in Fig.~1, but for the wavelength regime around 1268~nm. }
\end{figure}

\subsubsection{Telluric Spectrum}

The transmission spectrum of Earth (telluric spectrum) was calculated following the procedure outlined by Seifahrt et al. (2010). Like in the previous section, this procedure
uses the LBLRTM code developed by Clough (2005) and the transmission molecular absorption database HITRAN containing the 42 most 
prominent molecules and isotopes present in the atmosphere of Earth. As an input to LBLRTM, we adopted a model of the Earth's atmosphere that contains meteorological information for temperature and pressure as a function of height for an average night at a given observatory. In our case we retrieved the weather information from the Global Data Assimilation System (GDAS). GDAS models are available in 3 hour intervals for any location around the globe\footnote{GDAS webpage: {\tt http://ready.arl.noaa.gov/READYamet.php}}. By employing LBLRTM, we then calculated absorption and emission spectra for a given path through the atmosphere, i.e. for a given airmass. In our model we have assumed that the observations take place  at a ground-based observatory at an altitude of about 2500 meters above sea level, and we have accounted for a changing airmass values during the course of the night within the range $X=1.1$ to $2$. 
The transmission spectrum around the $O_2$ A-band and the $O_2$ band at 1268 nm for an airmass of 1.3 is illustrated as example in the middle panels of Figures 1 and 2.  We note that a few H$_2$O absorption lines appear in the telluric spectrum around 1290~nm. These lines partly overlap with the O$_2$ lines and constitute an additional source of noise. In our simulations, we have considered the depth of these lines to be constant for simplicity reasons. In fact, however, the depths of the water absorption lines may strongly vary during the course of a night (e.g. Rodler et al. 2012), an effect that should be implemented in future simulations.  

\subsubsection{Stellar Spectra}

To model the emission from the star, we focus on M-type stars with spectral types M1V--M9V. The reason for this, as already explained by Kaltenegger \& Traub (2009)
and Snellen et al. (2013), is that the
planetary transit depths are inversely proportional to the squared radius of the star and therefore, the transmission spectrum signal of an exo-Earth will be easier to detect 
around a star with smaller radius.  For a star like the Sun (G2V), the ratio of the areas between the stellar disk and the atmosphere of an exo-Earth's ring is of the order of 5$\times$ $10^5$, while for an M1V star that ratio is $\sim$ 1.25$\times$$10^5$, and for an M9V star the ratio drops to $\sim$ 4$\times$$10^3$.  

We adopted stellar model spectra for effective temperatures and surface pressures corresponding to M1V--M9V stars using the new library of high-resolution spectra based on the PHOENIX code and presented in  Husser et al. (2013). This new library has been generated using a new equation of state, which improves the treatment of chemical equilibrium in stellar atmospheres, and updated atomic and molecular line lists. These new models provide a better match to observed spectra of M dwarfs than the latest PHOENIX (Brott et al. 2005) and ATLAS9 (Castelli et al. 2003) models. For simplicity, we assumed solar abundances in all the models. As an example, we show the model spectrum generated for an M4V star with effective temperature $T_{\rm eff} =3000$~K, gravity $\log g = 4.5$~dex and solar abundances in the bottom panels of Figures 1 and 2.

\subsubsection{Telescope+Instrument Setup}

There are three Extremely Large Telescopes (ELTs) with diameters 24 meters or larger expected to start operations in the next decade. The smallest of those 
ELTs is the Giant Magellan Telescope (GMT; www.gmto.org) with an effective diameter of 24.5 meters. One of the first light instruments on the GMT will be G-CLEF (Szentgyorgyi et al. 2012), which will be a high-resolution spectrograph with resolution modes between R = 25,000--120,000 and spectral coverage between 350--950 nm,  suitable to study the $O_2$ A-band. 

The second largest of the ELTs is the Thirty Meter Telescope (TMT; www.tmt.org). Among the TMT's expected suite of instruments the best suited to detect $O_2$ in the atmosphere of an Earth-like planet is the High Resolution Optical Spectrometer (HROS: Crampton et al. 2008), listed among the telescope's {\it first decade} instruments. 
HROS will provide a resolution of $R = 50~000$ for a 1 arcsec slit, or $R>90~000$ with an image slicer, and spectral coverage between $\sim$ 310--1000 nm. An image slicer is a device that allows to feed a large fraction of the stellar flux into a narrow slit, thereby avoiding slit losses and at the same time allowing to employ high spectral resolving powers (see e.g. Dekker et al. 2003).

The largest of the ELTs will be the European Extremely Large Telescope (E-ELT), with a diameter of 39-meters. In the case of the E-ELT, there are two spectrographs with resolution $R> 100~000$ being studied on Phase A\footnote{http://www.eso.org/sci/facilities/eelt/instrumentation/phaseA.html}, CODEX (Pasquini et al. 2010) and SIMPLE (Origlia et al. 2010). CODEX only covers wavelengths up to $\sim$ 710 nm, while the wavelength coverage of SIMPLE starts at $\sim$ 800 nm. Therefore none of these instruments include the $O_2$ A-band at 760 nm. SIMPLE, on the other hand, will cover the $O_2$ band at 1268~nm. A third high-resolution spectrograph called HIRES, is now being proposed as one of the first light instruments for the E-ELT, although the specifications for this instrument are not defined yet. 

\subsubsection{Noise Models}

The last of the simulation inputs is noise. The noise in an observed spectrum can be divided into random (white) and correlated (red) noise. While the distribution of white noise is Poissonian and can be reduced by increasing the S/N of the spectrum, the red noise is not random and cannot be reduced by increasing the S/N. However, the red noise can be related to instrumental and atmospheric effects, such as e.g. variations of airmass or changes of the position of the spectrum on the detector. Analysis techniques have improved in recent years to levels where it is now possible to significantly reduce the amount of red noise in the data. 

In our simulations we have added both sources of noise. White noise is simulated using a random distribution function with a maximum amplitude set by the S/N of the particular model (see section 2.2). In the case of red noise, its pattern can contain a series of frequencies that depend on the specific cause of the noise for a given dataset. Therefore, we have simulated red noise using the approach described in (Kasdin 1995), by assuming a functional form of 1/f, where $f$ is the frequency. The amplitude of the red noise is set to a fraction of the white noise in an specific model. For our simulations we consider red noise contributions of 0 to 100$\%$ of the white noise. Typical red noise levels of ground based data remain of the order of 20$\%$ of the white noise in the optical and 50-100$\%$ of the white noise in the near-infrared (see e.g. Pont et al. 2006, Rogers et al. 2009).

\subsection{Generation of Model Spectra}

Using the inputs described above we generate model spectra, $C$, using a expression of the form
\begin{equation} \label{E1}
C=\Big(a~(1+\epsilon^{-1})^{-1} T \Big) \otimes G~,
\end{equation} 
where
\begin{equation} \label{E2}
a=(1+v_\star c^{-1})~S + \left(1+(v_\star+v_{\rm pl})c^{-1} \right) \epsilon^{-1}~P~.
\end{equation} 

In these equations, $c$ denotes the speed of light, $S$ is the spectrum of the host star, $P$ is the transmission spectrum of the planetary atmosphere, and $T$ is the telluric spectrum of Earth for a given airmass.  $G$ is the instrumental profile (IP) of the spectrograph. In our simulations, we assume the IP to be a Gaussian function that degrades the spectral resolution of the combined spectrum to a chosen value. The remaining parameters in the equations are $v_\star$, the relative stellar velocity with respect to Earth (i.e. systemic velocity and barycentric velocity), $v_{\rm pl}$, the instantaneous radial velocity of the planet with respect to its host star (which is close to 0~km~s$^{-1}$ at a primary transit),  and $\epsilon$, the area ratio between the stellar disk and the atmospheric ring of the planet.


Once we have computed the spectrum $C$, we interpolate it onto a pixel grid, which simulates a given instrumental layout. Since the exact parameters of most instruments described in section 2.1.4 are not yet fully defined, we adopt two different pixel grid scales; one with a velocity resolution of 0.75~km~s$^{-1}$~pixel$^{-1}$ and the other 1.2~km~s$^{-1}$~pixel$^{-1}$. The 0.75~km~s$^{-1}$~pixel$^{-1}$ grid scale corresponds to the planned scale of G-CLEF for resolutions $R>25,000$ (Szentgyorgyi et al. 2012), while the 1.2~km~s$^{-1}$~pixel$^{-1}$ grid scale corresponds to UVES (Dekker et al. 2000), which is currently mounted at the 8.2 Very Large Telescopes (VLT) of the European Southern Observatory (ESO).

The next step is to adjust the flux of the model spectra based on the expected flux of the exo-Earth's host star that is received on Earth. Given that the potential transiting exo-Earth host stars will be at a range of possible distances from Earth, and therefore will have different apparent magnitudes, we consider a range of brightnesses for stars of a given spectral type. (Snellen et al. 2013) and (Kaltenegger et al. 2009) computed the most likely apparent magnitude of a nearby transiting exo-Earth host star based on the solar neighbourhood mass function and transit probability of an exo-Earth and they provided detectability estimations that are based on that number (see e.~g.~Section 3 in  Snellen et al. 2013). Here we adopt a different approach and instead of making any assumption about the apparent magnitude of the transiting exo-Earth host star, we assume that such star can be at any distance within 20 pc. Based on the star's distance, we adjust its apparent magnitude between 1 and 20~pc \footnote{We start our models at 1 pc based in the assumption that there is no other star closer to the Sun than Proxima Centauri at 1.33 pc. The 20~pc upper limit is chosen because of the faintness of M-dwarfs. In the remaining paper we only consider distances up to 11 pc since exo-Earths around stars further than that distance take too long to observe (see sections 3 and 4).}. The effective temperature and radius of the stars adopted for each spectral type are listed in Table 1. The table also shows the values of the area ratio between the stellar ring and the atmospheric ring of the planet, $\epsilon$, the I-band and J-band magnitudes of each star at 10 pc, the orbital period of a planet in the middle of the habitable zone of the star based on the definition by Kasting (1993), and the duration of the transits of such planet.

The flux that we finally measure from a star of a given magnitude will depend on the telescope and instrument throughput. Therefore, we estimated those fluxes using online Exposure Time Calculators (ETC) for each instrument, whenever available. For all ETCs, we assumed a median seeing of 0.8" and airmass values ranging from 1.1 to 2.0. 
G-CLEF already has an ETC\footnote{G-CLEF ETC version 1.0; http://alerce.astro.puc.cl/gclef.html}, while  for SIMPLE, we adopted the throughput value estimates listed on the project webpage\footnote{SIMPLE http://simple.bo.astro.it/}. The instrumental layout of HROS has not been published yet; we therefore assumed that HROS will be similar to G-CLEF, with a factor of 1.5 shorter integration times due to the larger diameter of the TMT with respect to the GMT. For a spectrograph with a UVES-like layout, we used the UVES ETC\footnote{ETC version 3.2.13; Website: http://www.eso.org/observing} and scaled the results accordingly to the diameter ratio between the ELT and the VLT with 8.2~m. For this instrument, we calculated the S/Ns for different slit widths, resulting in different spectral resolving powers. In addition, we also determined the S/N when employing the UVES image slicer \#3, which has a resolution of  $R=110,000$ (Dekker et al. 2003).

With the ETCs we estimated the exposure time needed to attain the maximum number of counts for each star's apparent magnitude, but at the same time avoiding reaching typical non-linearity regimes of 36~000~ADU and 12~000~ADU (in the brightest pixel on the chip) in a single exposure, respectively for the CCDs and near-infrared detectors. We furthermore accounted for the integration time and the dead time between two exposures to estimate the observations duty cycle for each star. While for CCDs we assumed a dead time of 15 seconds and a read-out noise of 5~electrons per exposure, we accounted for a dead time of 10 seconds and a read-out noise of 10~electrons for infrared detectors. The maximum single exposure time was chosen to be 600~seconds, since at longer exposures the radial velocity shift of the remote planet would smear out the absorption lines in its atmospheric spectrum.

Combining the duty cycle of each star with the flux collected per integration we obtained the total number of photons collected during the duration of one transit (see last column in Table 1). The duty cycles for each star, for each instrument, are summarized in Tables 2 and 3. The remaining columns in these tables are explained in section~3.4.

Finally, we added Poisson noise and red noise to each model based on their flux levels (i.e. S/N), following the description in section 2.1.5. Before adding the noise, we scaled each spectrum in such a way that the actual S/N per spectral pixel corresponded to the square root of the flux at that pixel.  This approach ensures that regions of telluric lines as well as stellar absorption lines have a higher noise and a lower S/N than the continuum. 



\subsection{Data sets and their analysis}\label{S22}

We generated data sets simulating transmission spectra of planetary transits using eqs.~1 and 2 with the following range of parameters, $v_\star = -150$ to  $150$~km~s$^{-1}$, $\epsilon = 4\times10^3 - 5.2\times10^5$, depending on the spectral type of the host star, and 
spectral resolutions of $R = 60,000 - 110,000$,  as well as a set of S/N levels including red noise contributions from 0 to 100$\%$  of the white noise level. In each data set, we simulated the observations during a night. Hence, each data set consisted of a series of spectra, which differed from each other by the airmass and the instantaneous radial velocity shift  of the planetary transmission spectrum ($v_{\rm p}$) with respect to its host star during transit. Additionally, we factored in the effects of the change of the barycentric velocity of the star plus its planet during the course of a night, which is of the order of a few times 100~m~s$^{-1}$.  Furthermore, to rule out selection effects coming from the adopted pixel grid, we also varied the zero-point of the pixel grid up to the average velocity span of one pixel. The duration of the datasets corresponds to twice the duration of the transits listed in the last column of Table 1.

To analyze each transit dataset in search for the extraterrestrial signal of $O_2$, we followed the approach outlined in Rodler et al.~(2012, 2013) and removed both a model telluric spectrum and a model stellar spectrum from each synthetic spectrum, thereby calculating a residual spectrum which -- ideally -- only contained the transmission spectrum of the planet, plus noise. We then carried out cross-correlations between the residual spectrum and the transmission model spectrum of the exo-Earth, and determined the cross-correlation function. For all spectra of each data set, we then summed up the cross-correlation functions in the rest frame of the planet and determined the radial velocity value that yielded the highest correlation (candidate feature).

To determine the number of false positives (i.e. the confidence level of the candidate feature), we  employed a 
   bootstrap randomization method (e.g. Rodler et al. 2013).    
   Random values of the orbital phases were assigned to the observed spectra, thereby creating $N$ different data sets ($N=100,000$ in our analyses).
   Any signal present in the original data was then scrambled in these
   artificial data sets.
   For all these randomized data sets, we re-ran the data analysis in the given parameter search ranges and located the candidate feature with its maximum value.
   
The confidence level of the candidate features was estimated to be $\approx 1-{\rm FAP} = 1-m/N$, where FAP is the false alarm probability and $m$ is the number of the best fit models  having a peak value larger or equal than the peak value of the cross-correlation function found in the original, unscrambled data sets. 

A detection is considered significant when the signal is at least $3\sigma$, which corresponds to a FAP of $\le 0.0027$.

\begin{deluxetable}{lccccccc}\label{T1}
\tabletypesize{\scriptsize}
\tablecaption{Host Star and Habitable Zone Planet Parameters}
\tablewidth{0pt}
\tablehead{
\colhead{spectral type} & 
\colhead{$T_{\rm eff}$ (K)} & 
\colhead{$R$ ($R_{\odot}$)} &  
\colhead{$\epsilon$} &  
\colhead{$M_I$ (mag)} &  
\colhead{$M_J$ (mag)} &  
\colhead{$P$ ~ (d)} &  
\colhead{transit duration (h)} 
}
\startdata
G2V & 5800 & 1 & 520~000      & 4.1  & 3.6 & 365.2 & 13.1  \\
M1V & 3600 & 0.49 & 125~000    & 7.7  & 6.4 & 43 &4.0     \\
M2V & 3400 & 0.44 & 101~000    & 8.3  & 6.5 & 33 & 3.4     \\
M3V & 3250 & 0.39 &  80~000    & 8.8  & 7.1 & 27 & 3.0     \\
M4V & 3100 & 0.26 & 35~000  &	10.0 & 7.9 & 16 & 2.1	\\
M5V & 2800 & 0.20 & 21~000  &	11.2 & 8.6  & 9.5  & 1.5	\\
M6V & 2600 & 0.15 & 12~000  &	 12.4& 10.1 & 6.0 & 1.1     \\
M7V & 2500 & 0.12 & 7~500   &	 13.6& 10.7 & 4.1 & 0.78	\\
M8V & 2400 & 0.10 & 6~000   &	 13.9& 11 & 3.3 & 0.69	\\
M9V & 2300 & 0.08 & 4~000   &	 14.7& 11.6 & 1.9 & 0.43	\\
\enddata
\end{deluxetable}

\begin{deluxetable}{l|cccc|cccc}\label{T2}
\tabletypesize{\scriptsize}
\tablecaption{Simulations for E-ELT spectrographs in the visual$\dagger$}
\tablewidth{0pt}

\tablehead{
~ & \multicolumn{4}{|c|}{UVES-like design} & \multicolumn{4}{c}{G-CLEF-like design}\\ \hline
\multicolumn{1}{c|}{sp. type} & 
\colhead{obs. time (h)} & 
\colhead{duty cycle} & 
\colhead{transits}   &
\multicolumn{1}{c|}{time (years)} & 
\colhead{obs. time (h)} & 
\colhead{duty cycle} & 
\colhead{transits}   &
\colhead{time (yrs)}  
}
\startdata
G2V &  840 &  0.05   & 65 & $\ge 65$    & 429	&  0.08 &   34  & $\ge 33$ \\
M1V  &  110 &  0.58   & 28 & $\sim 29$   &  67 &  0.71 & 17  &	$\sim18$ \\  
M2V  &  98 &  0.69   & 29 &  $\sim 23$  &  62 &  0.80 &  19 &	$\sim15$ \\  
M3V  &  88  &  0.79   & 29 &  $\sim 19$ &  57 &  0.87 & 19  &	$\sim13$   \\
M4V  &  43   &  0.92  & 21 &  $\sim 8$  &  29  &  0.95 & 14  &	$\sim6$ \\   
M5V  &  46   &  0.97  & 31& $\sim 7$    &  31  &  0.98 & 21  &	$\sim5$ \\   
M6V  &  43   &  0.98  & 39& $\sim 6$    &  30  &  0.98 & 27  &	$\sim4$ \\   
M7V  &  35   &  0.98  & 45& $\sim 5$     &  24  &  0.98 & 31  &  $\sim3$  \\  
M8V  &  34   &  0.98  & 48& $\sim 4$    &  25  &  0.98 & 36  &  $\sim3$   \\  
M9V  &  29   &  0.98  & 66 & $\sim 3$   &  22  &  0.98 & 50  &  $\sim2$   \\  
\enddata
\tablenotetext{\dagger}{For a velocity span of 1.2~km~s$^{-1}$~pixel$^{-1}$ (UVES-like design, left) and 0.75~km~s$^{-1}$~pixel$^{-1}$ (G-CLEF-like design, right). Numbers are given for 5 pc distance.}
\end{deluxetable}

\begin{deluxetable}{l|cccc|cccc}
\tabletypesize{\scriptsize}
\tablecaption{Simulations for G-CLEF @ GMT and HROS @ TMT}\label{T3}
\tablewidth{0pt}
\tablehead{
~ & \multicolumn{4}{|c|}{ G-CLEF @ GMT} & \multicolumn{4}{c}{HROS @ TMT}\\ \hline
\multicolumn{1}{c|}{sp. type} & 
\colhead{obs. time (h)} & 
\colhead{duty cycle} & 
\colhead{transits}   &
\multicolumn{1}{c|}{time (years)} & 
\colhead{obs. time (h)} & 
\colhead{duty cycle} & 
\colhead{transits}   &
\colhead{time (yrs)}  
}
\startdata
G2V  & 470   &  0.18 &   37  & $\ge 37$ & 448  &  0.13 &   35  & $\ge 35$ \\
M1V   &  133 &  0.86 & 33  &   $\sim35$  &  97 &  0.81 & 24  & $\sim26$ \\
M2V  &  133 &  0.91 &  40 &   $\sim31$  &  95 &  0.87 &  28 &  $\sim23$ \\
M3V   &  130 &  0.94 & 44  &   $\sim28$  &  91 &  0.92 & 30  & $\sim20$   \\
M4V   &  70  &  0.98 & 34  &   $\sim14$  &  48  &  0.97 & 23  &   $\sim9$ \\
5V   &  79  &  0.98 & 53  &   $\sim12$  &  53  &  0.98 & 36  &   $\sim9$ \\
M6V   &  75  &  0.98 & 68  &   $\sim10$  &  51  &  0.98 & 46  &   $\sim7$ \\
M7V   &  61  &  0.98 & 78  &  $\sim8$	 &  41  &  0.98 & 53  &  $\sim6$  \\
M8V   &  69  &  0.98 & 100  &  $\sim8$   &  46  &  0.98 & 67  &  $\sim6$ \\
M9V   &  67  &  0.98 & 154  &  $\sim7$   &  45  &  0.98 & 104  &  $\sim5$ \\
\enddata
\\For a distance of 5 pc
\end{deluxetable}

\begin{deluxetable}{l|cccc}
\tabletypesize{\scriptsize}
\tablecaption{Simulations for SIMPLE @ E-ELT}\label{T4}
\tablewidth{0pt}
\tablehead{ Sp. type &  obs. time (h) & duty cycle & transits   &
time (yrs)}
\startdata
M3V  &  713 &  0.12 & 238 &  $\ge 100$ \\ 
M4V  &  173 &  0.41 & 82  &  $\sim 33$ \\   
M5V  &  99 &  0.69 & 66 & $\sim 15$  \\ 	  
M6V  &  38 &  0.75 & 35 & $\sim 5$   \\ 	  
M7V  &  21 &  0.84 & 27 & $\sim 3$   \\ 	  
M8V  &  18 &  0.87 & 26 & $\sim 2$   \\ 	  
M9V  &  10 &  0.91 & 24 & $\sim 1$   \\ 	  
\enddata
\\For a distance of 5 pc
\end{deluxetable}



\section{Results}

Using the simulations described above, we have investigated a series of questions pertaining the successful ground-based detection of $O_2$ in the atmosphere of an Earth twin around  a nearby star. Those questions include the optimal spectral resolution of the observations, the limitations imposed by the relative radial velocity of the star and Earth, the limitations imposed by the depth of the telluric $O_2$ lines, and the number of transits we need to observe to achieve at least a 3$\sigma$ detection in the presence of white and red noise. The answers to those questions are provided in Figures 3-8 and Tables 2-4, and are also described in detail in the text below.

The simulations were carried out for G-CLEF mounted on the 24m~GMT, for HROS on the TMT assuming an instrument configuration similar to G-CLEF, and for three instruments mounted on the E-ELT; the near-infrared spectrograph SIMPLE and two hypothetical high-resolution spectrographs in the visual,  a  G-CLEF-like instrument and UVES-like instrument.

\subsection{Spectral Resolution}

We used the UVES ETC to investigate the efficiency of observations with an UVES-like instrument at different spectral resolutions. We carried out simulations for resolutions
between 60,000 and 110,000, rescaling the S/N of the contimuun of each spectrum to the aperture of the E-ELT. Figure 3 shows the result of those tests for the particular case of an Earth twin orbiting around an M4V star at a distance of 5 pc. In this case, the optimal spectral resolution appears to be $R=80,000$, which corresponds to the best trade-off between slit losses and the spectral smearing of the exo-Earth's signal. The plot also shows, for comparison, the result of the same test when using an image slicer -- in this particular case we simulated image slicer \#3 described in Dekker et al. (2003). When using the image slicer, which is only used at resolution $R = 110,000$,  the efficiency of the observations improves by a factor of two with respect to the efficiency of the instrument at that same resolution when using only a slit. We note that this result represents the ideal case, i.e. we did not account for possible changes of the instrumental profile induced by an image slicer, which might hamper the data analysis.  The tests also show that the use of the image slicer improves the overall efficiency of the observations, where only about 21 transits are necessary to achieve a 3$\sigma$ detection of the exo-Earth's atmosphere, while 25 transits are necessary to achieve the same precision level when just observing at a spectral resolution of $R=80,000$ without an image slicer.

The result of these tests are instrument-specific and similar tests will have to be done for other instruments to determine the optimal resolution for this type of science once their final specifications are defined. However, we can argue that employing pre-slit optics, such as image slicers capable of concentrating a larger fraction of the stellar flux into the slit, can significantly reduce the number of transits needed for a successful detection, for all instruments.


\subsection{Relative Radial Velocity}

The detectability of an Earth twin depends, among other things, on how many of its $O_2$ lines can be observed. Our atmosphere produces the same absorption features (telluric lines), at the same wavelength positions (cf. Figs.~1 and 2), and in addition, the lines have some intrinsic width. Therefore, the Earth twin spectrum will be only detectable when its lines are Doppler shifted with respect to the telluric lines by certain amounts. To investigate this effect we generated a set of simulations in which the spectrum of the Earth twin was shifted with respect to the spectrum of Earth by values of the relative velocity between the host star and Earth between -150~km~s$^{-1}$  
and +150~km~s$^{-1}$, and determined the fraction of lines that would appear blended in each case. The results of those simulations are illustrated in Figure 4 for both the $O_2$ A-band and the $O_2$ band at 1268 nm. 

In the case of the $O_2$ A-band, we find that most lines will appear blended as long as the relative velocities are $<$ $\pm$ 15~km~s$^{-1}$. Line blending becomes significant again (with about 50$\%$ of the lines blended) at relative velocities of about $\pm$ 50~km~s$^{-1}$, coinciding with the average separation between individual lines in the $O_2$ A-band spectrum. The optimal relative velocity regimes (with 10$\%$ or less of the lines blended), appear to be -15 to -30~km~s$^{-1}$ and 15 to 30~km~s$^{-1}$. 

In the case of the $O_2$ band at 1268 nm, which are narrower than the A-band lines, we find that relative radial velocities in the regimes around -85 to -65~km~s$^{-1}$, -10 to 10~km~s$^{-1}$, and 65 to 85~km~s$^{-1}$ are also significantly blended. Stars with any other relative velocity value can be observed with the guarantee that no more than about 20$\%$ of the $O_2$ lines will be blended.


\begin{figure}\label{F3}
\includegraphics[angle=-90,scale=0.31]{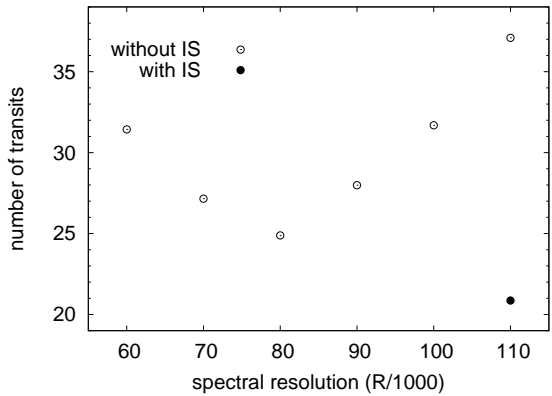}
\caption{Simulation results for the number of transit observations necessary to obtain a 3$\sigma$ detection of $O_2$ around 760 nm in the atmosphere of an Earth twin orbiting a M4V star at 5pc. The instrumental setup is an UVES-like instrument mounted on the 39-m E-ELT. The open circles show the result of the simulations for spectral resolutions between 60,000 and 110,000. The filled circle shows the result of the same simulations when an image slicer at a resolution of 110,000 is used. These simulations assume a relative radial velocity for the star of 20~km~s$^{-1}$ with respect to Earth and no red noise.}

\end{figure}

\begin{figure}\label{F4}
\includegraphics[scale=0.45]{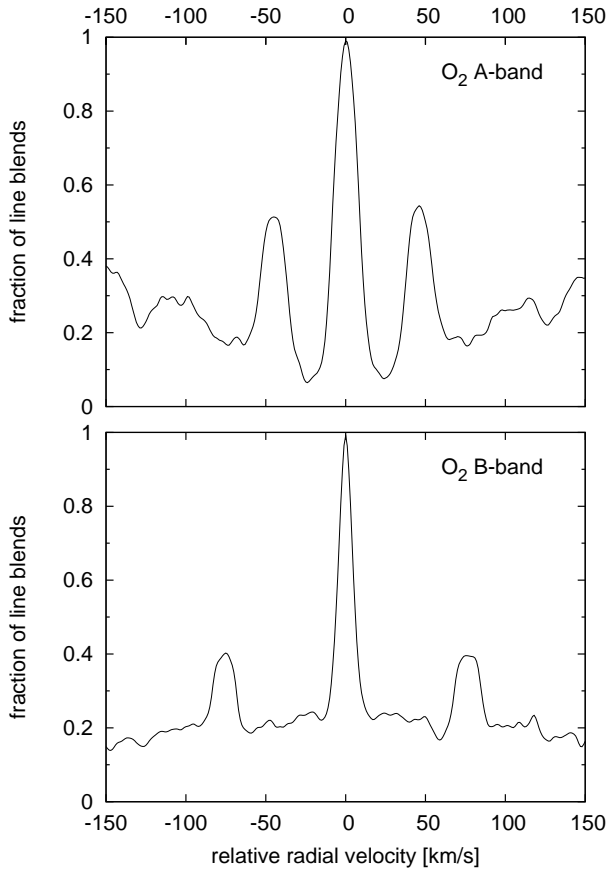}
\caption{Plot showing the normalised fraction of line blends between the telluric spectrum and the transmission spectrum of the Earth twin as a function of the relative radial velocity between the host star and Earth. The upper panel corresponds to the $O_2$ A-band at 760 nm and the bottom panel to the $O_2$ band at 1268 nm. Based on this plot, observations are best suited when the relative velocity of the star is between $\pm$(15 and 30)~km~s$^{-1}$ in the case of the A-band, and for velocities outside the ranges $\pm$(85 to 65)~km~s$^{-1}$ and -10 to 10~km~s$^{-1}$ in the case of the $O_2$ band at 1268 nm. }

\end{figure}

\begin{figure}\label{F5}
\includegraphics[angle=-90,scale=0.31]{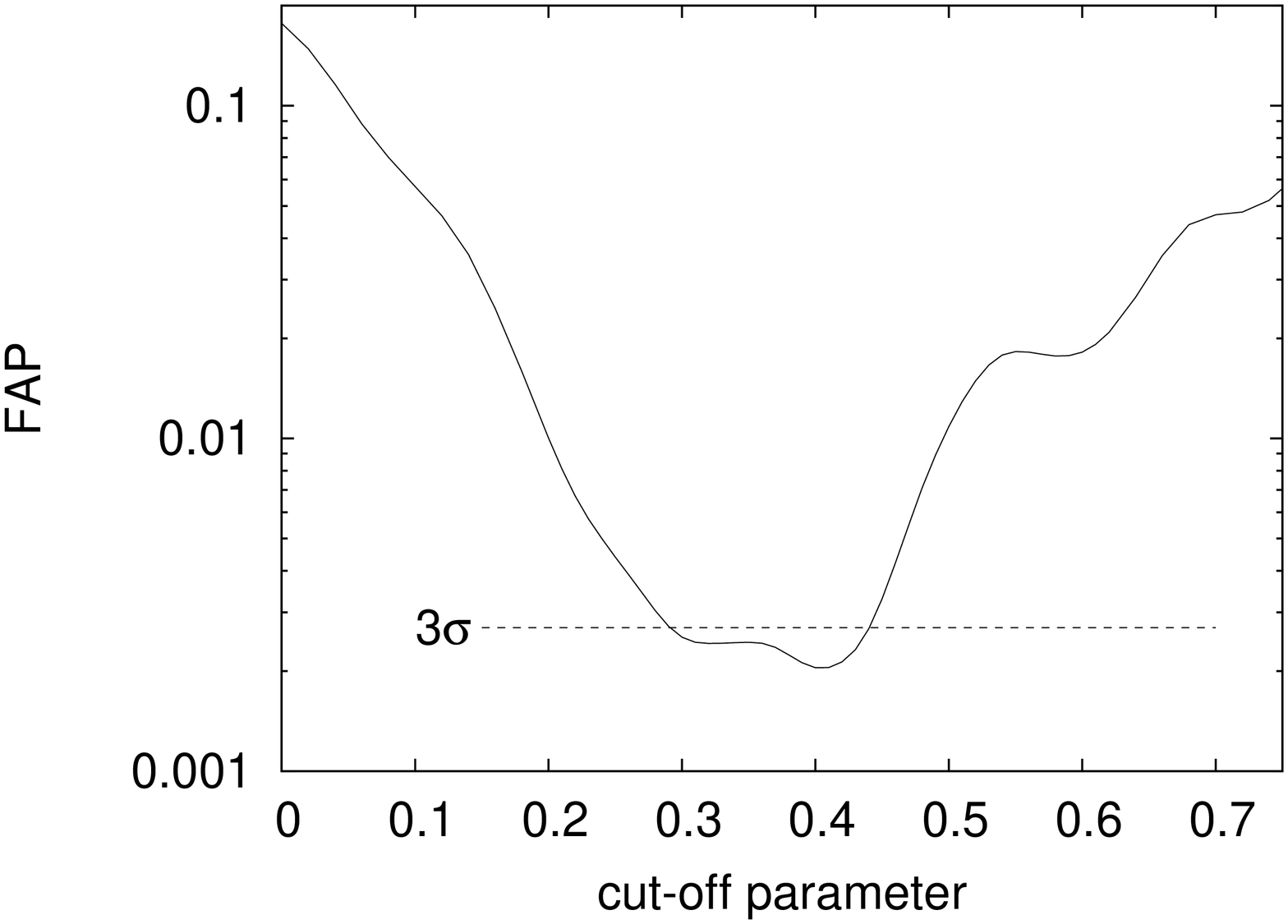}
\caption{False Alarm Probability (FAP) of the Earth twin's transmission spectral signal detection as a function of the cut-off parameter value. We find that values of the cut-off parameter between $\sim$ 0.29 and 0.43 give a FAP lower than 0.027, which corresponds to a 3$\sigma$ detection. The lowest value of the FAP corresponds to a cut-off parameter value of 0.4.}

\end{figure}

\subsection{Telluric Line Depths}

Another effect affecting the detectability of an Earth twin is noise introduced by telluric contamination. In other words, telluric lines block a large fraction of the host star's light, causing the S/N of its spectrum to drop significantly around those wavelengths. In the case of the $O_2$ band at 1268 nm, telluric contamination around 1268~nm causes the S/N ratio to drop to 0.3 with respect to the S/N ratio in the continuum. In the case of the $O_2$ A-band, telluric contamination around 760 nm fully blocks all incoming light at some wavelengths (see middle panels in Figures 1 and 2). The best strategy to avoid this source of noise during the data analysis process is to skip wavelength regions where the S/N of the observed spectrum drops below a certain value. Therefore, we define a cut-off parameter with respect to the S/N of the spectral continuum for which lines (or portions of the spectrum) to keep during the analysis. The optimum value of this cut-off parameter is a trade-off between keeping a large enough number of absorption features from the planet and avoiding noisy regions due to strong telluric contamination. Adopting a non-optimal cut-off parameter will result in strongly suppressing the exo-Earth's signal and might lead to a non-detection (i.e. to a detection with confidence level less then 3$\sigma$).  

To determine the value of this cut-off parameter we analyzed different datasets by masking out different wavelength regions in which the telluric absorption lines dropped below a certain amount, e.g.  a cut-off parameter of 0.5 corresponds to masking out all the wavelength regions for which the S/N drops by 50$\%$ with respect to that of the continuum. The result of these tests is shown in Figure 5, where we represent the FAP of the exo-Earth's detection as a function of the cut-off parameter in S/N units (i.e. a value of 1 for the cut-off parameter means that even the continuum is masked out). The figure indicates that the best value of the cut-off parameter is approximately 0.4 times the S/N of the continuum.


\subsection{Number of Transits}

The last question we addressed is how many transit observations are necessary to achieve at least a 3$\sigma$ detection of an Earth twin's atmosphere, and what will be the time span needed to collect those many transits. To estimate the number of transits needed, we focused first on the 760 nm $O_2$ A-band and assumed a transiting planetary system with an Earth twin in its habitable zone, at a distance of 5 parsecs from Earth and moving away from us with a relative velocity of 20~km~s$^{-1}$. We determined the duty cycle of the observations by assuming an instrument dead time of 15 sec after each exposure (see section 2.2). To calculate the number of transits, we divided the observing time by the transit duration of the planet as a function of stellar spectral type. The values for the orbital periods and transit durations were taken from Kaltenegger \& Traub (2009) and are provided in Table~1. To estimate the overall time span of the observations, we factored in the orbital period of the planet and assumed that only every 9th transit can be observed, after accounting for object visibility from Earth, transit occurrence, and the relative radial velocity between the target and Earth (see also Section~4). The results are summarised in Tables~2 and 3. 

Table~2 is split into two halves; the left-hand side lists the results for a spectrograph with a UVES-like layout (i.e. spectral resolving power of $R=110,000$, image slicer \#3, velocity span of 1.2~km~s$^{-1}$~pixel$^{-1}$) installed on the E-ELT, while the right-hand side shows the values determined for a spectrograph with a design similar to G-CLEF (i.e. $R=100,000$, velocity span of 0.75~km~s$^{-1}$~pixel$^{-1}$), also on the E-ELT.  G-CLEF's design seems to be significantly more efficient. 

In this table, and in Table~3, we also show the times needed for an Earth twin orbiting a G2V star. In this case, because of the very low duty cycle and the very shallow transit depth, it would take over 30 transits, and therefore more than 30 years to detect $O_2$ with 3$\sigma$ confidence. Once we also factor in for the number of observable transits from the ground, these observations would take many decades. For M1V - M3V stars we estimate the time span of the observations to be $\sim$ 13 - 19 years. For later type M-dwarfs, where the orbital periods of planets in their habitable zones are shorter, this number drops significantly to about 6 years for M4V stars and about 2 years for M9V stars. For a planet orbiting a M4V star, it would require a minimum of 14 transits  to detect $O_2$ with 3$\sigma$ confidence in its atmosphere. In the case of M9V stars, which are intrinsically fainter, it would take at least 50 transits. 

Figure~6 shows the number of transits required for a 3$\sigma$ detection with the two instrumental setups in Table~2, as a function of distance and for spectral types M3V through M7V. Planets around an M3V star need more transits at very low distances due to a decrease in the duty cycle of the observations, i.e. the stars become too bright and the observing times need to be short to avoid saturation. Therefore most of the observing time in those cases is spent on instrument readouts. The same applies for earlier type stars. The figure also shows how for stars located at distances larger than about 8-10 parsecs, the number of transits necessary to achieve a 3$\sigma$ detection becomes too large.

Table~3 is similar to Table~2, but showing the results for two other observational configurations; the left-hand side of the table shows the results of our simulations for G-CLEF mounted on the GMT. The right-hand side shows the result of the simulations for HROS mounted on the TMT.  For both spectrographs we assume a spectral resolution of $R = 100,000$ and a velocity span of 0.75~km~s$^{-1}$~pixel$^{-1}$. The top panel of Figure~7 shows the number of transits required for a 3$\sigma$ detection with G-CLEF on the GMT as a function of distance and, as in Figure~6, for spectral types M3V through M7V. As expected, in this case, the number of transits necessary to achieve a detection is larger, by approximately a factor of two, than the number of transits necessary with an instrument like G-CLEF in the larger aperture E-ELT (also see bottom panel of Figure~6). The number of transits in Figure~7 decrease by a factor of about 1.5 for the case of HROS mounted on the TMT. 

Finally, Table~4 summarizes the results of our simulations for the near-IR spectrograph SIMPLE to be mounted on the E-ELT. In this case the simulations focus on the 1268 nm $O_2$ band, while we still assume a transiting planetary system with an Earth twin in its habitable zone, at a distance of 5 parsecs from Earth and moving away from us with a relative velocity of 20~km~s$^{-1}$. We adopted a spectral resolution of $R = 100,000$, a velocity span of 1.2~km~s$^{-1}$~pixel$^{-1}$ and a dead time between two subsequent exposures of 10 seconds. We only carried out simulations for stars with spectral types later than M3V, since the number of transits necessary for an M3V star was already over 200. The results on that table reveal that using the $O_2$ band at 1268 nm to detect $O_2$ in the atmosphere of an Earth twin will require a significantly larger amount of telescope time than observations of the $O_2$ A-band for all spectral types earlier than M7V. The bottom panel of Figure~7 shows he number of transits required or a 3$\sigma$ detection with SIMPLE  as a function of distance and for spectral types M5V through M9V. In this case, the number of transits decreases steadily with increasing spectral type.

\begin{figure}\label{F6}
\includegraphics[scale=0.45]{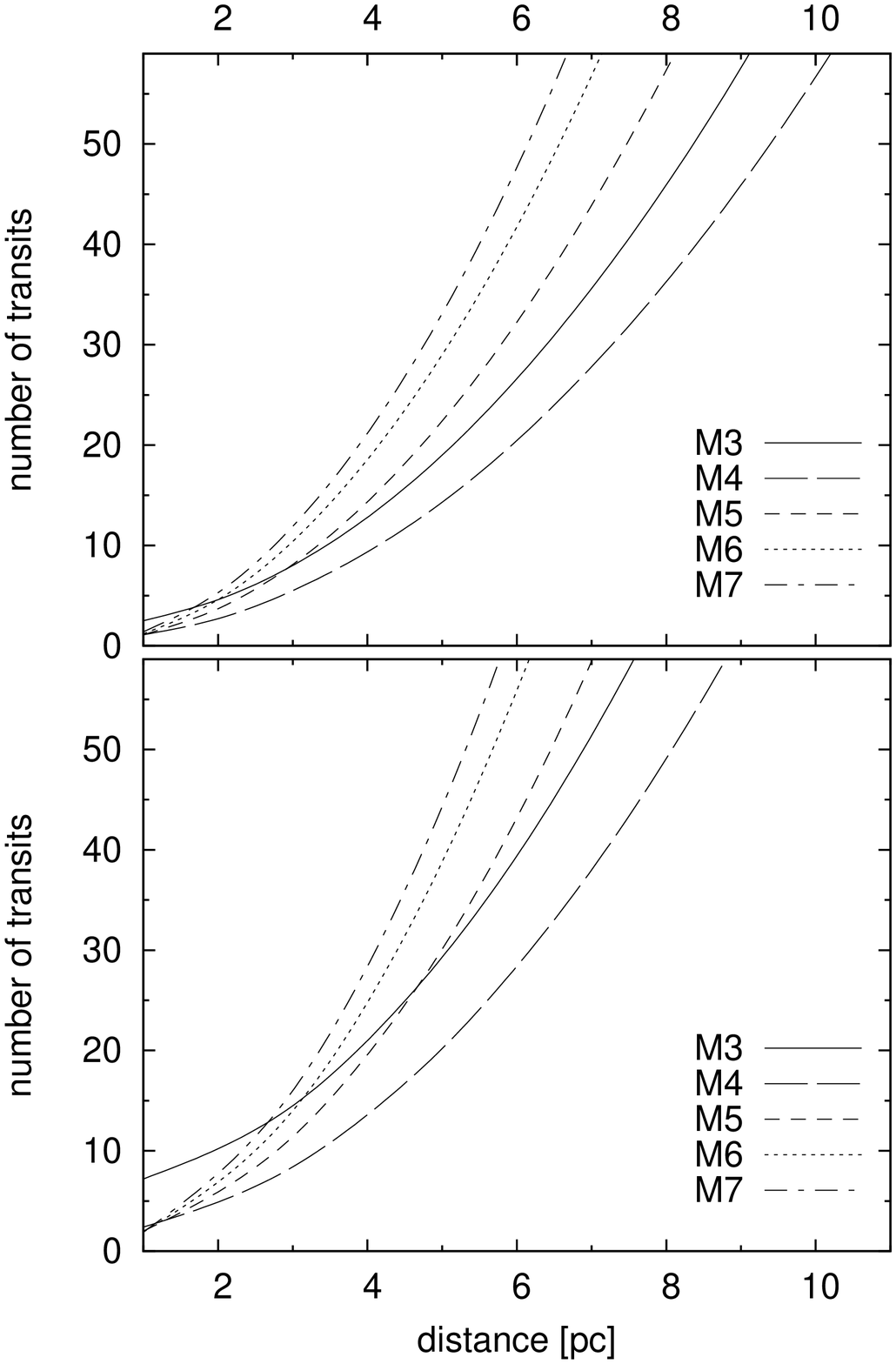}
\caption{Number of transits required to measure the 760~nm $O_2$ absorption band with $3\sigma$ confidence  versus distance of the planetary system from Earth. As instrument, we assume a high-resolution spectrograph mounted on the 39-m~E-ELT. The upper panel illustrates the results for a G-CLEF-like instrumental layout which provides a velocity span of 0.75~km~s$^{-1}$~pixel$^{-1}$ and a spectral resolution of $100,000$. The lower panel depicts the results for a UVES-like spectrograph with velocity span of 1.2~km~s$^{-1}$~pixel$^{-1}$, an image slicer and a spectral resolution of $110,000$.
 In these plots, we assume no red noise. To account for different red noise levels, the number of transits needs to be multiplied by the values provided in Figure~8.}
\end{figure}

\begin{figure}\label{F78}
\includegraphics[scale=0.45]{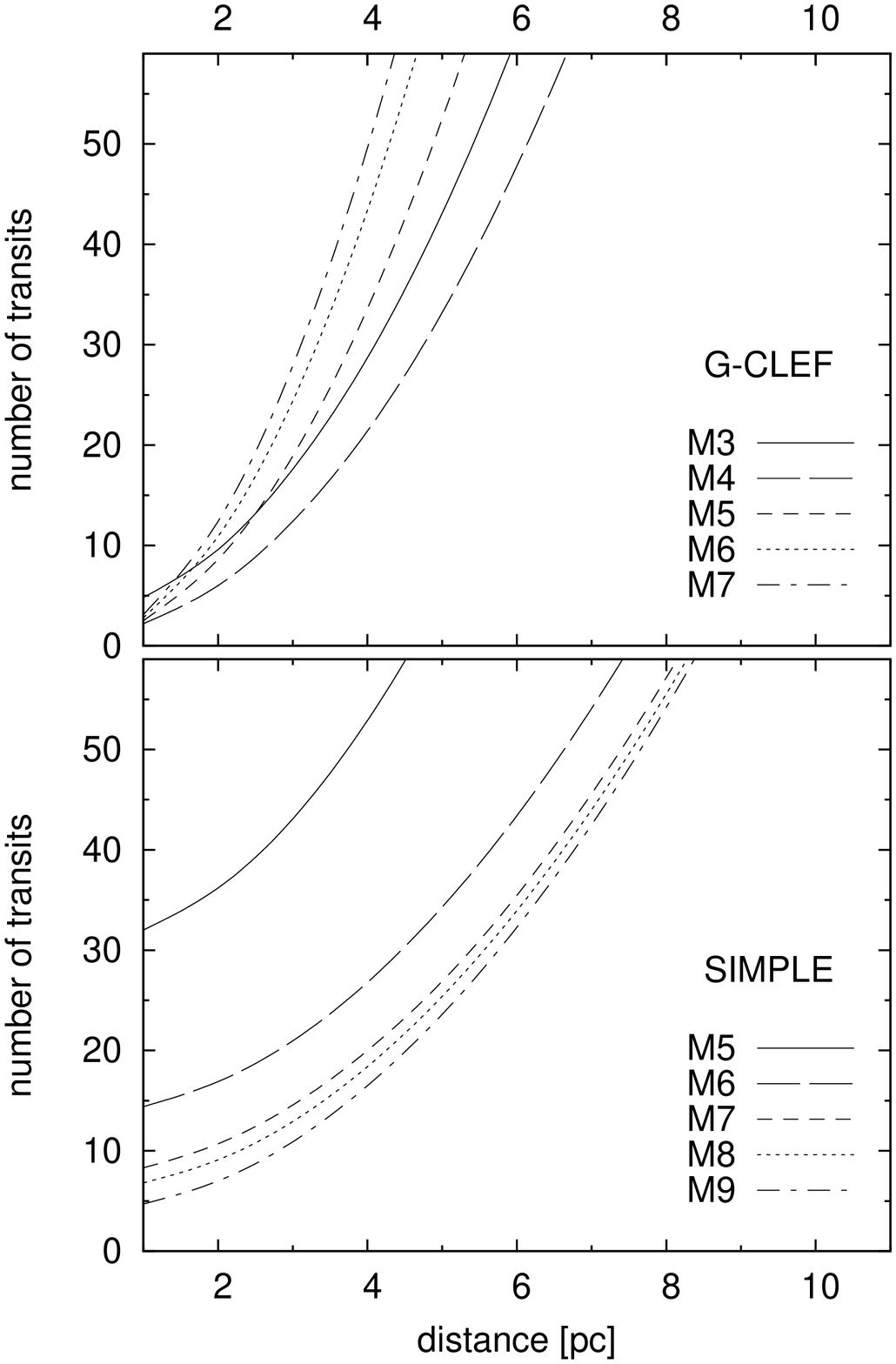}
\caption{{\it Top panel:} Number of transits required to measure the 760~nm oxygen absorption band with G-CLEF with $3\sigma$ confidence  versus distance of the planetary system from Earth.  {\it Bottom panel:} Same as above, but for measurements of the 1268~nm oxygen absorption band with the SIMPLE-spectrograph employing a spectral resolving power of $100,000$. In these plots we assume no red noise. To account for different red noise levels, the number of transits needs to be multiplied by the values provided in Figure~8.}
\end{figure}

\subsubsection{The Effect of Red Noise}

Tables~2,~3, and~4,  as well as Figures~6 and~7 show the results for the ideal case, i.e. without any contributions from correlated (red) noise. However, typical red noise levels are 20\% and 50\%  of the white noise level for data recorded in the visual and near-infrared, respectively (Pont et al. 2006, Rogers et al. 2009). The higher red noise level in the near-infrared wavelength regime is due telluric contamination and to a larger extent, due to detector noise. To make the simulations more realistic, we added different levels of red noise to the simulated data sets and analysed them as described before. Figure~8 shows the difference in the amount of observing time required to achieve a 3$\sigma$ detection as the red noise levels increase. For a red noise level of 20\%  of the white noise level, the observing time required to attain a $3\sigma$ detection rises by a factor of $\sim1.4$. This means that for a red noise level of 20\%  of the white noise level, the observing times as well as the number of transits shown in Tables 2 and 3 and in Figure~6 and~7 (top panel) needs to be multiplied by this value. This effect becomes more severe for higher red noise levels. e.g. for correlated noise levels of 50\% and 100\%  of the white noise level, the time factor increases by $\sim2.6$ and $\sim9$, respectively.

When comparing the results determined by accounting for typical red noise levels (20\%  of the white noise level in the visual, 50\%  of the white noise level at near-infrared wavelengths), we find that the lowest time span of observations is achieved with a G-CLEF-like spectrograph mounted on the E-ELT for all M-dwarfs with spectral types M1V-M8V. Observations in the $O_2$ band at 1268 nm are more efficient than in the visual only for M9V stars (see Figure~9).


\begin{figure}\label{F7a}
\includegraphics[angle=-90,scale=0.31]{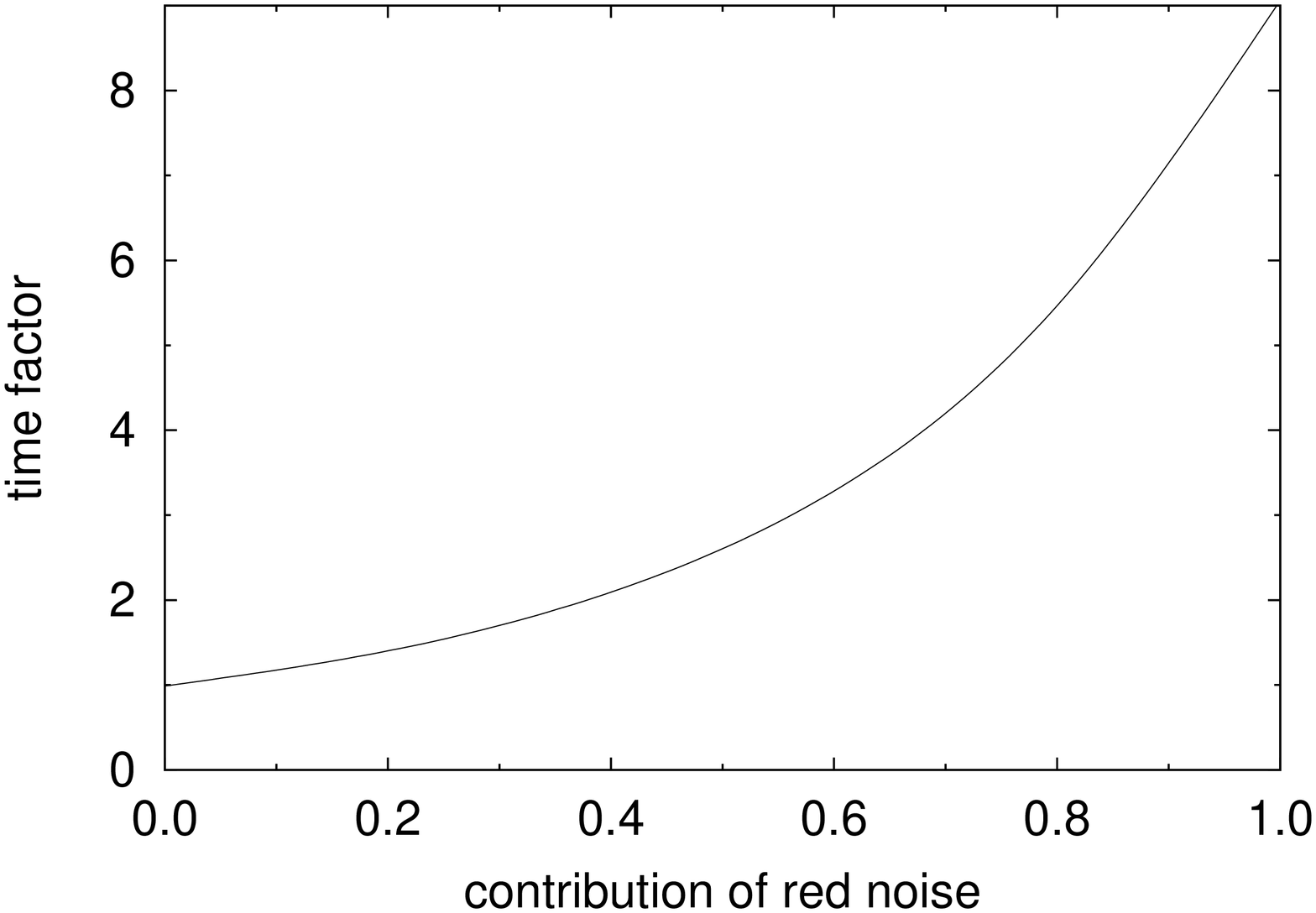}
\caption{The effect of red noise in the number of transits and the amount of time necessary to detect $O_2$ in the atmosphere of an Earth twin at a 3$\sigma$ confidence level. For example, a red noise level of 40$\%$  of the white noise level in the observed spectra approximately doubles the observing time required for a detection.}
\end{figure}
\begin{figure}\label{F9}
\includegraphics[scale=0.31, angle=-90]{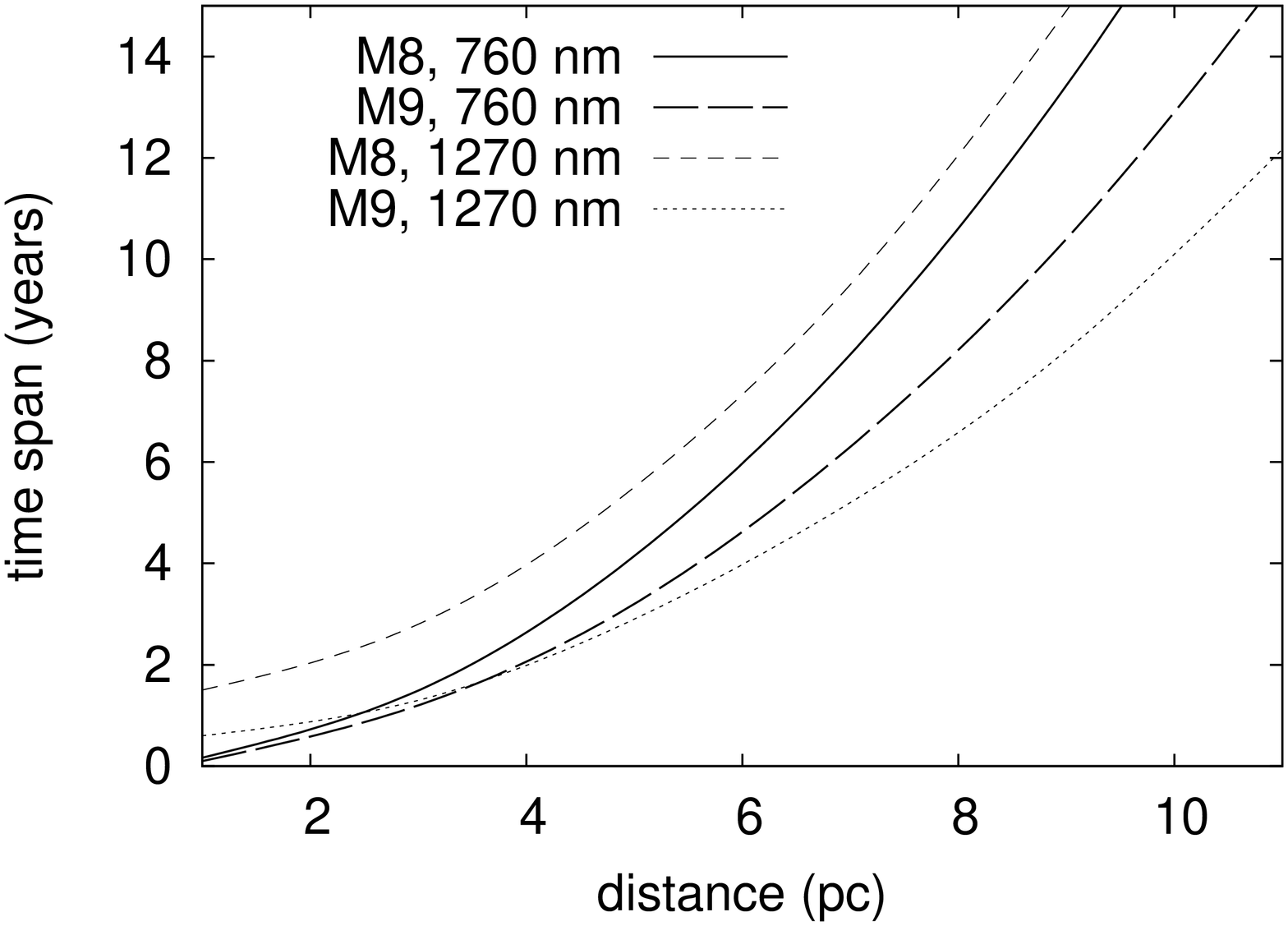}
\caption{Time span of the observations versus distance of the host star for observations of M8V and M9V stars in the $O_2$ A-band at 760 nm and the $O_2$ band at 1268 nm. The observations at 760 nm have a red noise level of 20$\%$, while the observations at 1268 nm have a red noise level of 50$\%$  of the white noise level. Unlike in the ideal case of no red noise, the observations in the $O_2$ band at 1268 nm are more efficient than the observations on the $O_2$ A-band only for M9V stars. These results are based on SIMPLE and a G-CLEF-like instrument mounted on the E-ELT.}
\end{figure}

\section{Discussion \& Conclusions}

We have carried out detailed feasibility studies to probe molecular oxygen in the atmosphere of an Earth-analogue orbiting M-dwarfs with future instruments to be mounted on upcoming ELTs. In our simulations, we factored in the stellar spectra for different spectral types, the telluric spectrum of the Earth, different instrumental settings of possible future instrumentation, refraction effects in the Earth analog's atmosphere, and a variety of noise models, including the effects of correlated noise (red noise) in the data. When we compare our work to previous, similar studies, these are the most detailed simulations to date. 

\begin{figure}\label{F11}
\includegraphics[scale=0.31,angle=-90]{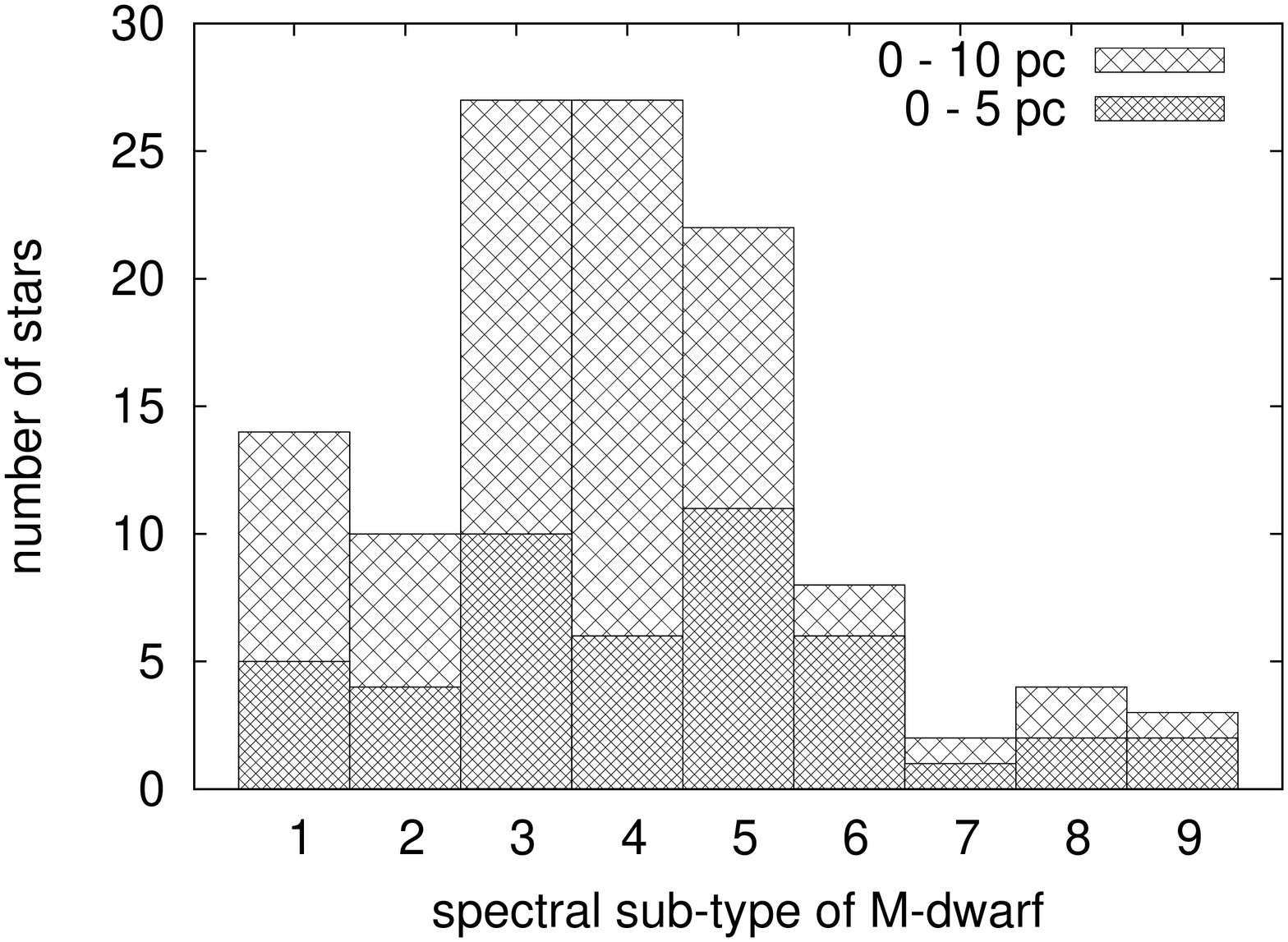}
\caption{Frequency of M-dwarfs in the solar neighbourhood (distances $d\le10$~pc) as a function of spectral type. The most common sub-types are M3 and M4 dwarfs. For distances up to 5~pc, the most frequent sub-types are M3 and M5 dwarfs. }
\end{figure}

\begin{figure}\label{F10}
\includegraphics[scale=0.39]{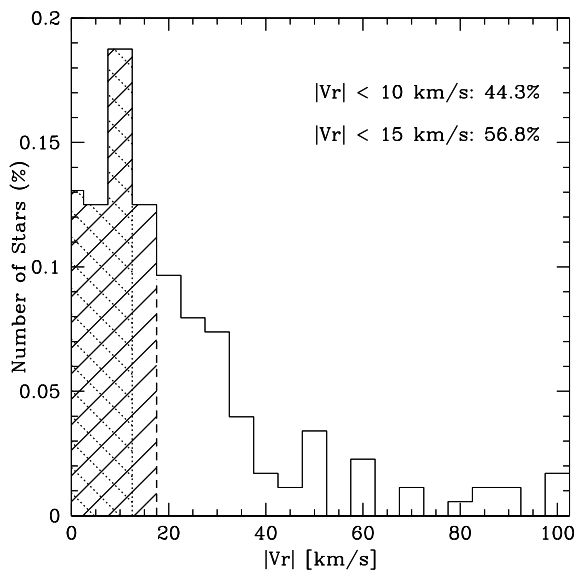}
\caption{Histogram of the distribution of barycentric radial velocities for a sample of 176 M-dwarfs published by Chubak et al. (2012) in 5~km~s$^{-1}$ bins.  . The bins of stars with velocities less than 15 and 10~km~s$^{-1}$, respectively, are shown as dashed and square shaded areas in the histogram. Assuming the Chubak et al. (2012) sample is representative of all nearby M-dwarfs, about 44.3$\%$ of the stars will have absolute radial velocities smaller than 10~km~s$^{-1}$, and 56.8$\%$ will have radial velocities smaller than 15~km~s$^{-1}$.}
\end{figure}

\begin{figure}\label{F12}
\includegraphics[scale=0.31,angle=-90]{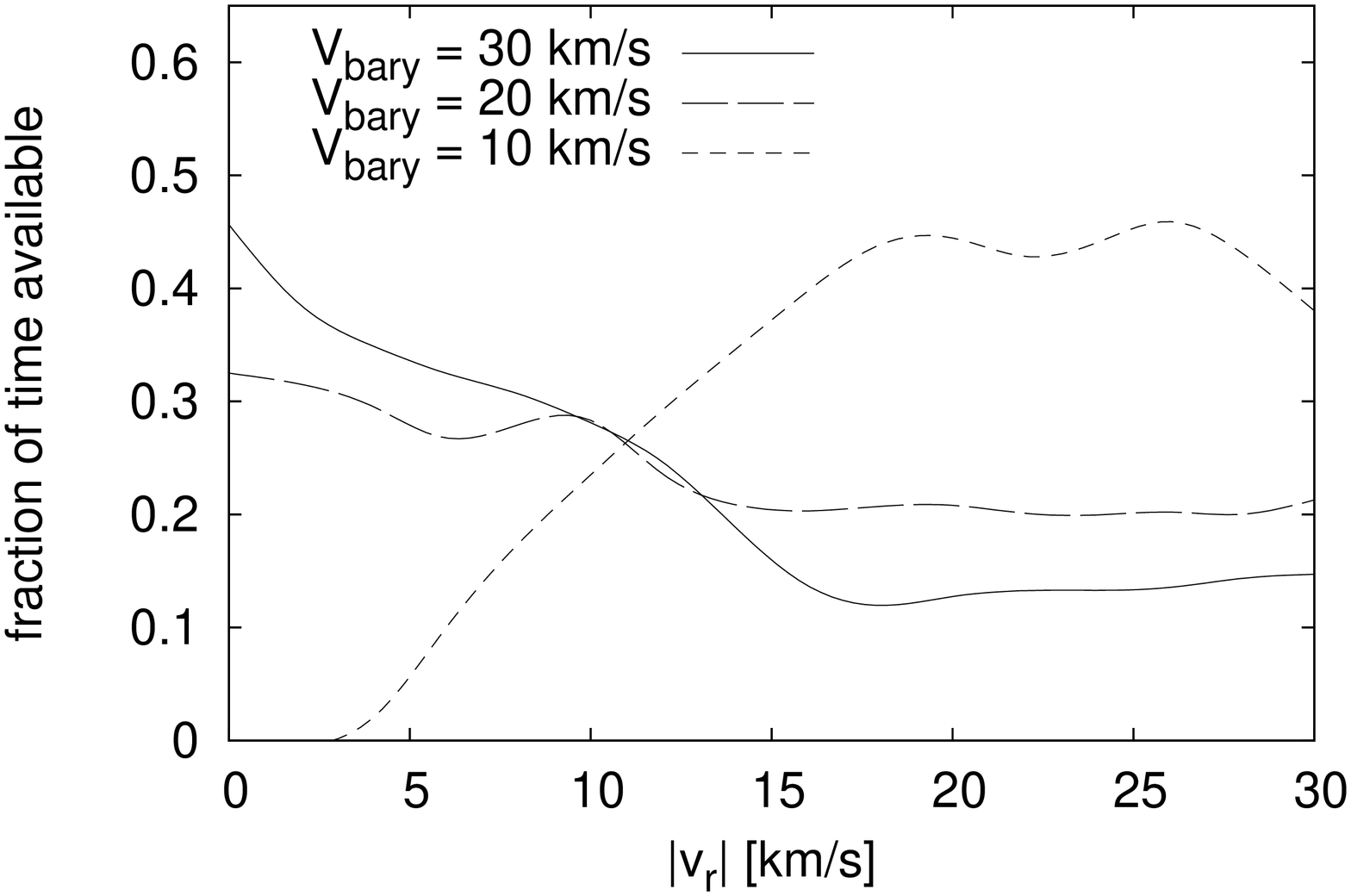}
\caption{Amount of time of optimum visibility of an object in units of the yearly observing time available at an ideal observing site as a function of the absolute value of the systemic radial velocity $|v_r|$. We only consider the target to be visible when following criteria are fulfilled: (a) air mass $X\le2$, (b) Sun $15^\circ$ below horizon, and (c) a relative radial velocity of $\pm$(15 to 30)~km~s$^{-1}$.
We note that the vast majority of stars exhibit semi-amplitudes of the barycentric velocity of $V_{\rm bary}>20$~km~s$^{-1}$.
}
\end{figure}

A result of our investigation is that a G-CLEF-like instrument mounted on an E-ELT is the best suited for the detection of $O_2$ in the atmosphere of an Earth-like planet orbiting M1V-M8V stars. Only for planets orbiting M9V stars, observations carried out in the near-infrared become more efficient than in the visual. This is a result of lower red noise levels at visible wavelengths, but the situation could change in the future if infrared red noise levels can be improved to match noise levels in the visible. In that case, near-infrared observations will be more efficient for planets around M7V stars and smaller.

Fig.~10 depicts the frequency of M1V to M9V stars in the solar neighborhood\footnote{Data were retrieved from the RECONS database (http://www.recons.org).}.  M3 and M4 dwarfs are the most common sub-types, with more than 25 stars of each spectral type within a distance of 10~pc. When combining this distribution with the results of our simulations, we conclude that  very close-by M4V stars will be the most likely candidates for the detection of oxygen in the atmosphere of an Earth-like planet. In just 40~hours of observing time ($\sim 20$ transits) a 3$\sigma$ detection of an object at 5~pc from Earth could be attained with a G-CLEF-like instrument mounted on the 39~m E-ELT. For this estimation, we assumed a red noise level of 20\% of the white noise level. A typical orbital period of a planet located in the habitable zone around an M4-dwarf is 16~days. Once the object visibility, orbital period and the relative radial velocity are taken into account, we estimate that it would nevertheless take a total time span of observations of $\sim 8$~years to accumulate sufficient data for a detection. This number soars up to  175~hours (74 transits) and a time span of roughly 30 years for an object at a distance of 10~pc, which is the most probable distance to find an Earth-analogue around such a star (Kaltenegger \& Traub 2009).

We compare our results to previous studies by other groups. Snellen et al. (2013) carried out simulations with the goal to detect the $O_2$ A-band in the atmosphere of Earth-analogues orbiting around M4V and M6V stars, with apparent magnitudes $I=10$ and $I=11.8$, respectively. According to their results, it takes about 4 to 15 transits to attain a $3\sigma$ detection of $O_2$. They assumed a line contrast ratio between stellar continuum and the deepest absorption lines of $20,000$, which corresponds to $\epsilon\approx35,000$ in the case of a spectral resolution $R=100,000$, which causes a line depth of 0.6 for the deepest lines with respect to the stellar continuum (c.f. line depths in Fig.~1, top panel). Furthermore, they assumed that the transit duration was 1.4~hours for planets orbiting those types of dwarfs. Adopting the values for $\epsilon$ and the transit duration listed in Table~1, our studies reveal that it takes a total of about 42 and 60 transits for a $3\sigma$ detection of oxygen in the atmosphere of an Earth-like planet orbiting an M4V and M6V stars, respectively. These results are for the ideal case, i.e. without red noise and a G-CLEF-like spectrograph mounted on the 39m~E-ELT.  If we include a typical 20\% red noise contribution in these studies, the numbers rise to 60 and 84 transits (i.e. 12-25 years), respectively for M4 and M6 dwarfs. Our results are far less optimistic than those presented by Snellen et al. (2013). 

Ideally,  the incoming flux from the planet host star should find its way to the data analysis. To record high-resolution spectra of a star, it is necessary to carry out the observations through a narrow slit (or fiber). This, however, means that only a fraction of the stellar light is used -- and in case of bad seeing conditions, it might happen that more than 60\% of the incoming light is lost. To avoid such an unfavorable situation, pre-slit optics should be employed which enable to feed more than 90\% of the photons of the star into the narrow slit. This can be achieved with image slicers or multiple fibers, although in this last case there is some concerns that transmission differences between fibers might affect the results. Such potential problems will have to be evaluated before hand.  Since the designs of most of the high-resolution spectrographs to be mounted on the ELTs  haven't been fixed yet, we strongly recommend to include pre-slit optics to collect most of the stellar light and feed it into the spectrograph. Based on our simulations in Figure~3, the usage of pre-slit optics could improve the required number of observations by a factor of two. In this estimation, however, we have not accounted for possible irregularities in the IP induced by an image slicer, which might represent a further noise source.

At this point, we want to emphasize that our simulations represent a somewhat ideal case, i.e. stable instrument configurations and perfectly fitting models in the data analysis.  Our simulations  need to be redone once the configuration details of the spectrographs are known.


For the planning of the observations, we find that a very important parameter to take into account is the relative radial velocity of the star with respect to the Earth. That relative radial velocity is the combination of the systemic radial velocity of the star ($v_r$) and the instantaneous barycentric velocity $v_{\rm bary}$. Our atmosphere produces the same line pattern that we search for in the atmosphere of a remote planet. At some wavelengths, the telluric contamination almost absorbs all the incoming flux, leading to very low S/N ratios in those wavelength regions, which need to be masked out in the data analysis. It is therefore crucial to minimize the numbers of absorption line blends by only taking observation when the telluric spectrum and the spectrum of the planet are Doppler shifted by a certain amount. For observations in the visual, we find that the optimum velocity range of the planet host star with respect to the Earth is limited to two small velocity windows, ranging from $\pm$(15 to 30)~km~s$^{-1}$.

In Figure~11 we show the distribution of absolute values of the barycentric radial velocities for a sample of 176 M-dwarfs reported by Chubak et al. (2012). Assuming that the distribution of radial velocities in that sample is representative of all nearby M-dwarfs, we can conclude that the majority of them have barycentric radial velocities between -15 and 15~km~s$^{-1}$. Therefore, what is the amount of time of optimum visibility of those objects, i.e. when they show relative radial velocities of $\pm$(15 to 30)~km~s$^{-1}$? To answer this question, we need to account for the barycentric velocity of the Earth, which causes a periodical radial velocity shift with a semi-amplitude $V_{\rm bary}$ of approximately 29.8~km~s$^{-1}$ for an object located close to the ecliptic plane. However, the semi-amplitude of the barycentric velocity is a function of the ecliptic latitude $\beta$ (i.e. the angular distance of the object from the orbital plane): 
\begin{equation} \label{E4}
V_{\rm bary} = (29.8 \cos \frac{\beta \pi}{180})~{\rm km~s}^{-1}.
\end{equation} 
For the vast majority of stars, the semi-amplitude of the barycentric velocity $V_{\rm bary}$ is between 20 and 29.8~km~s$^{-1}$.

Figure~12 shows the amount of time of optimum visibility of an object in units of the yearly observing time available at an ideal observing site. For this estimation, we assumed that the object can be observed at an airmass $X\le2$ between 2 hours and 8 hours (at opposition with Earth). Furthermore, for simplicity we assumed a circular orbit of Earth and factored in the instantaneous value of the barycentric velocity of the object. Figure~12 depicts that for semi-amplitudes of the barycentric velocity of $V_{\rm bary}\ge20$~km~s$^{-1}$, the atmospheric lines from the planet would be detectable a fraction of 0.2 to 0.4 of the time for the majority of the M-dwarfs, which have instrinsic radial velocities between -15 and 15~km~s$^{-1}$.

To understand possible error sources in the data analysis, we briefly review the main steps: We remove both the telluric spectrum as well as the stellar spectrum from the data, and finally attempt to retrieve the planetary signature from the residual spectrum via cross-correlation. The removal of both spectra is an iterative process, which involves the determination of the IP and the convolution of the stellar + telluric model spectra with the IP (the method is outlined in Rodler et al. 2013). While the telluric spectrum can be well reproduced with theoretical model spectra, the modeling of M-dwarf spectra is still in its infancy. In principle, if the instrument is stable during one night, it would be sufficient to take spectra of the star for some hours before and after the transit, subtract the telluric lines, then co-add these spectra to form one high-S/N spectrum for that night, which is then subtracted from the observed spectrum taken during the transit (cf. procedure used by Rodler et al. 2008). 
However, this procedure would drastically boost the need of observing time. Detailed studies on the modeling of M-dwarf spectra are clearly required.

Additionally, it is crucial to discard strong telluric features in the data analysis.  We want to make future observers aware  that regions which show a S/N ratio less than about 0.4 times the S/N ratio in the continuum of the spectrum should be masked out. Adopting a wrong cut-off parameter causes the the weak planetary signal to be suppressed. 

Another possible error noise constitutes the model spectrum of the planet atmosphere, which is adopted in the cross-correlation. We are attempting to detect a very weak planetary signal of the planetary atmosphere; the lack of some absorption features in the theoretical model would suppress the planetary signal in the cross-correlation function. To avoid this situation, a large numbers of atmospheric models with different mixing ratios and molecular configurations need to be probed.


We conclude that successful measurements of $O_2$ in Earth-like atmospheres with future ground-based instrumentation will be most likely limited to host stars of spectral types later than M3V, which are located at distances less than 8~pc. Our simulations demonstrate, that a $3\sigma$ discovery of $O_2$ in the atmosphere of such a planet would require even tenths of transits and several years of observing time at ELTs.




\acknowledgments
We are grateful to the constructive comments of the anonymous referee, who helped us to improve the manuscript.
FR acknowledges financial support from the Spanish Ministry of
Economy and Competitiveness (MINECO) and the "Fondo Europeo
de Desarrollo Regional" (FEDER) through grant AYA2012-39612-C03-01. This research has made use of NASA's Astrophysics Data System Bibliographic Services.

\end{document}